\documentclass[conference,9pt]{IEEEtran}
\IEEEoverridecommandlockouts
\usepackage{cite}
\usepackage{amsmath,amssymb,amsfonts}
\usepackage{algorithmic}
\usepackage{graphicx}
\usepackage{textcomp}
\usepackage{xcolor}

\usepackage{balance}
\usepackage{fixltx2e}
\usepackage{enumitem}
\usepackage{booktabs}
\usepackage{multirow}
\usepackage{multicol,makecell}
\usepackage{xspace}
\usepackage{graphicx}
\usepackage{xcolor}
\usepackage{fancyhdr}

\usepackage[normalem]{ulem}
\usepackage[bottom]{footmisc}
\usepackage{courier}
\usepackage{dblfloatfix}
\usepackage{longfbox}
\usepackage{caption}
\usepackage[]{hyperref}
\usepackage[colorinlistoftodos,prependcaption,textsize=scriptsize]{todonotes} % For margin notes

\usepackage{subcaption}

\usepackage{kotex}

\usepackage{xurl}

\usepackage{array}

\usepackage{tabularx}
\usepackage{bm}

\newcommand{\ms}[1]{{\color{black}{#1}}}

\newcommand{\jy}[1]{{\color{black}{#1}}}

\newcommand{\okj}[1]{{\color{black}{#1}}}

\newcommand{\head}[1]{{\noindent\textbf{#1.}\xspace}}

\newcommand{\degree}{\ensuremath{^\circ}C\xspace}

\newcommand{\vth}{V$_{\text{th}}$\xspace}
\newcommand\inum[1]{(#1)\xspace}
\newcommand{\vref}{V$_{REF}$\xspace}
\newcommand{\tailcut}{{\textsc{{TailCut}}}\xspace}
\DeclareRobustCommand\bcirc[1]{\tikz[baseline=(char.base)]{
           \node[shape=circle,draw,inner sep=0pt,fill=black, text=white] (char) {#1};}}

\newcommand{\STAR}{{\textsc{STAR}}\xspace} % 소문자 섞인 대문자 스타일
\newcommand{\baseline}{{\texttt{Baseline}}\xspace}
\newcommand{\StarSSD}{{\texttt{StarSSD}}\xspace}
\newcommand{\TcSSD}{{\texttt{TcSSD}}\xspace}
\newcommand{\stitle}{\underline{ST}ate-\underline{A}ware \underline{R}andomizer\xspace} % 밑줄 있는 버전
\def\BibTeX{{\rm B\kern-.05em{\sc i\kern-.025em b}\kern-.08em
    T\kern-.1667em\lower.7ex\hbox{E}\kern-.125emX}}

\title{STAR: Improving Lifetime and Performance of High-Capacity Modern SSDs Using State-Aware Randomizer}

\author{
    Omin Kwon$^{*1}$, 
    Kyungjun Oh$^{*1}$, 
    Jaeyong Lee$^{1}$, 
    Myungsuk Kim$^{2}$,
    Jihong Kim$^{1}$% 
    \thanks{$^*$These authors contributed equally to this work.} \\
    $^{1}$Seoul National University, Seoul, Republic of Korea \\
    $^{2}$Kyungpook National University, Daegu, Republic of Korea \\
}

\begin{document}

\maketitle

\begin{abstract}
Although NAND flash memory has achieved continuous capacity improvements via advanced 3D stacking and multi-level cell technologies, these innovations introduce new reliability challenges, particularly lateral charge spreading (LCS), absent in low-capacity 2D flash memory. 
Since LCS significantly increases retention errors over time, addressing this problem is essential to ensure the lifetime of modern SSDs employing high-capacity 3D flash memory. 
In this paper, we propose a novel data randomizer, \stitle (\STAR), which proactively eliminates the majority of weak data patterns responsible for retention errors caused by LCS.
Unlike existing techniques that target only specific worst-case patterns, \STAR effectively removes a broad spectrum of weak patterns, significantly enhancing reliability against LCS. By employing several optimization schemes, \STAR can be efficiently integrated into the existing I/O datapath of an SSD controller with negligible timing overhead.
To evaluate the proposed \STAR scheme, we developed a \STAR-aware SSD emulator based on characterization results from 160 real 3D NAND flash chips. 
Experimental results demonstrate that \STAR improves SSD lifetime by up to 2.3× and reduces read latency by an average of 50\%  on real-world traces compared to conventional SSDs.
\end{abstract}

\begin{IEEEkeywords}
3D NAND Flash memory, Multi-Level Cell, LCS, Reliability, Data Randomizer, NAND Controller
\end{IEEEkeywords}

\section{Introduction}
\label{sec:intro}

For decades, NAND flash memory has been a predominant technology to meet the storage capacity requirements of data-intensive applications such as graph analytics~\cite{liu-fast-2017, elyasi-fast-2019, matam-isca-2019} and machine learning-based applications~\cite{choe-arxiv-2017, liang-atc-2019}.
Due to key innovations such as 3D cell stacking and advanced multi-level cell technologies, NAND flash memory succeeded in continuous growth of flash capacity; the capacity of recent 3D triple-level cell (TLC) or quad-level cell (QLC) flash memory has increased up to 2 Tib per die~\cite{khakifirooz-sscl-2023, kim-isscc-2022}.
Specifically, 3D NAND flash memory~\cite{ishiduki2009optimal, jang2009vertical, park2014world} achieved high flash capacity by increasingly stacking flash memory cells in the vertical direction.
Furthermore, multi-level cell technology enabled flash memory to store multiple bits within a single flash cell, thus significantly reducing bit cost.

Such capacity-centric flash memory solutions have successfully provided high storage capacity, but conversely, they have intensified the reliability problems in modern SSDs.
It is well known that multi-level cell flash memory has disadvantages in flash reliability~\cite{cai-date-2012, cai-dsn-2015,cai-hpca-2017} due to its reduced error margin.
For instance, the lifetime of a QLC-based SSD, whose capacity is 2 times that of an MLC-based SSD, is only 30\% of the MLC-based SSD~\cite{amy-atc-2019,li-micro-2020,frickey-irps-2024}.
More seriously, 3D flash memory introduces new reliability problems that were not present in low-capacity 2D flash memory, called {\it lateral charge spreading} (LCS)~\cite{luo-acm-2018,lee2022tailcut}.
This unique reliability problem originated from the fact that commercial 3D flash memories adopt charge trap (CT)-type cell structures.
The CT-type flash cell exploits a non-conductive charge trap layer of silicon nitride (SiN) that traps electrical charges to store bit information, and all flash cells in a NAND string share a cylindrical-shaped trap layer that is physically connected vertically (Details in Section ~\ref{sec:background})~\cite{micheloni20163d, goda}.
When charge spreading happens, the trapped charge in a flash cell leaks out and moves to its (vertically) adjacent flash cells through the trap layer, which is similar to the diffusion mechanism.
Since the charge spreading makes bit information stored in a flash cell increasingly change over time (e.g., from `0’ to `1’), it is responsible for substantial retention errors in 3D flash memory.
Therefore, to ensure SSD lifetime, it is a critical challenge to address the LCS problem in high-capacity 3D flash memory (e.g., 3D TLC or QLC flash memory).

Although a large body of previous work has investigated the LCS spreading problem, most have focused only on device-level mechanisms of why LCS occurs between adjacent flash cells and under what conditions it is exacerbated~\cite{mizoguchi,luo-acm-2018,zhang-tvlsi-2022}. 
In the system-level approach, prior work has proposed a technique called TailCut, which attempted to reduce error-prone patterns by modifying encoded flash states~\cite{lee2022tailcut}. 
However, its effect can be limited in high-capacity 3D flash memory for two reasons.
First, since \tailcut requires redesigning the existing write/read operation, \tailcut can eliminate only the top 2 weak patterns due to its implementation complexity.
Second, to flip data bits, \tailcut should track bit information for all pages in the adjacent WLs. 
Therefore, such overhead makes \tailcut impractical in the advanced flash memory with four or more pages (e.g., QLC or PLC flash memory~\cite{khakifirooz-sscl-2023, lee-isscc-2024}).

Our goal in this work is to improve the lifetime and performance of modern high-capacity SSDs by mitigating the LCS problem.
To address the challenge, we propose a novel system-level optimization technique called \stitle (\STAR), which modifies the data randomizer in the SSD controller based on a thorough and systematic characterization of real 3D TLC/QLC NAND flash chips.
The key idea of \STAR is to selectively perform bit-flip operations for error-prone \vth states (e.g., P0, P1, P14, and P15 states in QLC flash memory), enabling the reduction of \emph{all} weak patterns vulnerable to the LCS problem.
% Unlike the existing approach that requires a high implementation overhead, \STAR attempts a minor change to the conventional LFSR-based randomizer with no modification to NAND flash chips, thus allowing our technique to work well in any type of flash memory and storage system.
\okj{Unlike existing approaches that require high implementation overhead, \STAR introduces a minor change to the conventional Linear Feedback Shift Register (LFSR)-based randomizer, which uniformly randomize data without considering error characteristics. In addition, \STAR does not require any modifications to NAND flash chips, so it can be applied to any type of flash memory and storage system.}

The \STAR operates in three sequential stages: \inum{i}~LFSR randomization, \inum{ii}~group error estimation, and \inum{iii} optimal bit-flip. 
First, \STAR scrambles the input data and distributes it uniformly across WLs and BLs using the conventional LFSR-based randomizer. 
Second, for each group of cells, \STAR estimates the error probability of a group and calculates the change in error probability if each bit-flip operation is applied.
% Bit-flip operations involve selectively inverting different combinations of physical bits in each cell (e.g., 16 bit-flip operations in QLC flash memory).
% By calculating the cumulative error change for each bit-flip operation based on our characterized state-specific error rates, \STAR identifies which operation would most effectively reduce the group's overall error probability.
\okj{Bit-flip operations selectively invert different combinations of physical bits in each cell (e.g., 16 options in QLC).
Using our profiled state-specific error rates, \STAR computes the total error change for each option and selects the one that best reduces group error.}
Finally, \STAR applies the optimal bit-flip operation to all cells in the group and records the corresponding bit-flip information, referred to as the Flip Indicator Bit (FIB), in the spare area of the target flash page.
This approach systematically reshapes \vth state distributions to favor lower-error states, statistically reducing weak patterns without explicitly detecting them. 
In subsequent read operations for the corresponding flash pages, the FIB information is used to reverse the bit-flip operations before performing de-randomization, ensuring data recovery.

\begin{figure}[t]
  \centering
  \setlength{\belowcaptionskip}{1pt}
  % \centering
  \includegraphics[width=\linewidth]{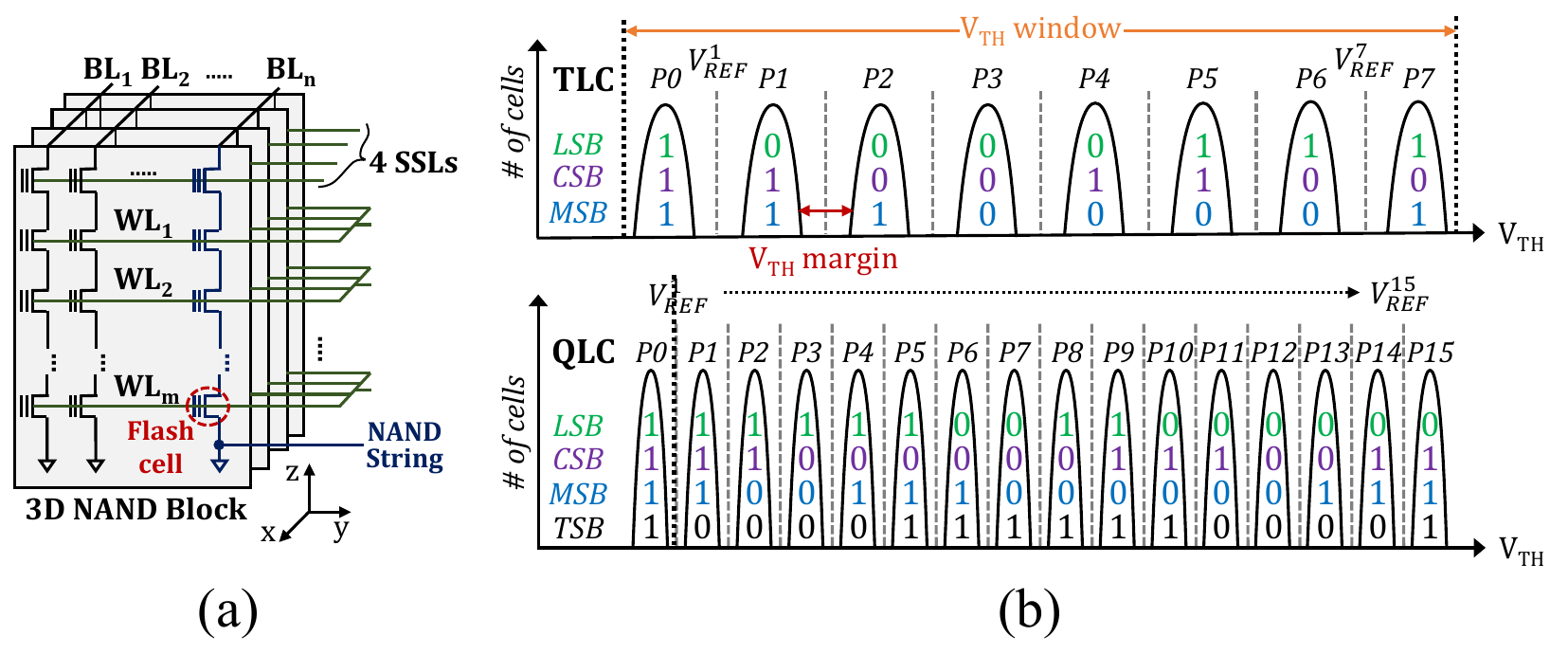}
  \captionof{figure}{(a) Organization of 3D flash memory. \jy{(b) Comparison of \vth distributions between TLC flash memory (top) and QLC flash memory (bottom).}}
  \vspace{-10pt}
  \label{fig:background1}
\end{figure}

Although the core idea behind \STAR is straightforward, implementing it in modern SSDs introduces several practical challenges.
One significant challenge is finding the optimal bit-flip operation without impacting I/O performance.
To address this, \STAR integrates three optimization techniques: (1) Zig-Zag IO Scheduling, (2) Group-Level Pipelining, and (3) Parallel Error Estimator.
The key approach is to execute various computational tasks in \STAR concurrently.
Zig-Zag IO scheduling interleaves the pages at the group level, making all physical bits (e.g., LSB, CSB, MSB, and TSB) in a group input to the \STAR at the same time, enabling them to process through \STAR's sequential stages together. Group-Level Pipelining of \STAR efficiently overlaps operations across different groups, enhancing the throughput efficiency of \STAR.  
In addition, the Parallel Error Estimator enables \STAR to evaluate the impact of all possible bit-flip operations on group error simultaneously.
Due to these latency-efficient optimizations, \STAR can fully exploit its reliability benefits while minimizing implementation overhead.

From extensive real-device characterizations, we validate the effectiveness of our proposed technique, \STAR.
In QLC flash memory, \STAR reduces the number of error-prone states (P0, P1, P14, P15) by 33.0\%, 30.1\%, 41.6\%, and 38.6\%, respectively, compared to conventional data randomization. 
Such redistribution of \vth states is directly converted to a substantial reduction in weak patterns created by a combination of error-prone states, which makes the top 10 weak patterns decrease by 71.7\%, on average.
The evaluation results reveal that \STAR improves the flash lifetime by up to 98\% (for TLC flash memory) and 130\% (for QLC flash memory), respectively, while 80\% (for TLC flash memory) and 50\% (for QLC flash memory) in the existing technique (e.g., \tailcut).
It strongly suggests that \STAR outperforms the existing scheme, and its impact on the flash lifetime is more pronounced in high-capacity QLC flash memory.

To evaluate the effectiveness of \STAR at the system level, we implemented a \STAR-enabled SSD, called \StarSSD, based on a state-of-the-art SSD emulator~\cite{kim2023virt}.
The evaluation results with various real-world traces demonstrate that \STAR significantly improves SSD lifetime up to 2.3x over the existing scheme. 
Furthermore, our technique efficiently reduces read latency by an average of 49\% due to fewer read retries compared to conventional SSDs.

\section{Background}
\label{sec:background}

\subsection{3D NAND Flash Memory Basics}
\ms{
\head{Organization}
Fig.~\ref{fig:background1}(a) illustrates the hierarchical organization of 3D flash memory.
Compared to 2D flash memory, 3D flash memory significantly increases its storage density by vertically stacking flash cells along the $z$-axis.
In our example, a 3D flash block consists of \emph{N} horizontal layers (h-layers) stacked along the z-axis.
Similarly, the 3D flash block may be described as having four vertical layers (v-layers) in the x-axis where each v-layer consists of \emph{N} stacked WLs separated by four select-line (SSLs)~\cite{micheloni20163d}.
A set of flash cells form a \emph{NAND string} that is connected to a bitline~(BL), and multiple BLs (e.g., 8KB-16KB) compose a \emph{block}.
The control gate of each cell at the same vertical location in a block is connected to a wordline (WL), so all the cells sharing the same WL concurrently operate.

\head{Multi-level Cell Flash Memory}
A flash cell stores bit data as a function of its threshold voltage (\vth) level, which highly depends on the amount of charge in the cell's charge trap layer; the more electrons in the charge trap layer, the higher the cell's \vth level.
It means that different bit data can be encoded by different \vth of the flash cell.
For example, we can assign the data value `0’ to a higher \vth and `1' to a lower \vth.
To increase storage capacity, a flash cell can store multiple bits by adjusting its \vth level more precisely, called \emph{m-bit multi-level cell} technology.
It stores $m$ bits within a single flash cell by using 2$^{m}$ distinct \vth states (i.e., $m$ is 3 and 4 for TLC and QLC flash memory, respectively).
The multi-level cell (MLC) technology was initially developed for storing 2 bits per cell~\cite{kanda201219,park2014three}, but it was extended to support QLC (4 bits/cell) and PLC (5 bits/cell).
Fig.~\ref{fig:background1}(b) illustrates the \vth distribution of a QLC flash memory that uses different $2^4$ states to store 4 bits per cell, indicating each WL contains four pages called LSB, CSB, MSB, and TSB page.

\head{NAND Flash Operation}
There are three basic operations that enable access to flash memory: program, erase, and read.
The program operation, which increases \vth of selected flash cells, injects electrons from the substrate into the charge trap layer of the selected flash cells using FN tunneling~\cite{fntunnel} by applying a high voltage ($> 20$V) to WL gates (i.e., changing the state of the flash cell from `1’ state to `0’ state).
On the other hand, to erase programmed cells (i.e., change the state of the flash cell from `0’ state to `1’ state), a high voltage ($> 20$V) is applied to the substrate (while WL gates are set to 0V) to remove electrons from the charge trap layer, which decreases the \vth of the flash cells.
The stored data can be read out by sensing the \vth level of the target flash cells on the selected WL using a proper read reference voltage (\vref).

\subsection{NAND Flash Reliability}
NAND flash memory is highly error-prone due to its imperfect physical characteristics, such as defects in the tunnel oxide layer of flash cells.
When a flash cell experiences repeated program and erase operations (e.g., P/E cycles), the high voltage used in program/erase operations physically damages flash cells.
Due to damage, the \vth distribution of cell gets wider and distorted, making the cells more error-prone, called \emph{endurance}.
In addition, a flash cell leaks its charge over time (i.e., its \vth level shifts to the left), which is called \emph{retention} loss.
If a cell's \vth level shifts beyond the \vref values (i.e., to adjacent \vth states having different bit values) due to endurance or retention, reading the cell causes a bit error.

Such a bit error occurs more seriously in high-capacity $m$-bit multi-level cell flash memory.
As $m$ is increased to store more bits within a flash cell, more \vth states should be put into the limited \vth window.
Furthermore, a \vth margin i.e., a gap between two neighboring \vth states calculated as $W_\text{Total}-\sum_{i=0}^{k} W_\text{Pi}$)) inevitably becomes narrower as shown in Fig.~\ref{fig:background1}(b).
When the \vth margin gets smaller, since two neighboring \vth states are more likely to be overlapped, NAND flash memory becomes more vulnerable to various noise effects, in turn, significantly increasing the raw bit-error rate (RBER) of NAND flash memory.
Therefore, to ensure data reliability, more careful reliability management is required for a higher $m$-bit multi-level cell flash memory (e.g., TLC or QLC flash memory).

\subsection{Mechanism of LCS}
\label{subsec:back_LCS}

\begin{figure}[t]
  \centering
  \includegraphics[width=0.85\linewidth]{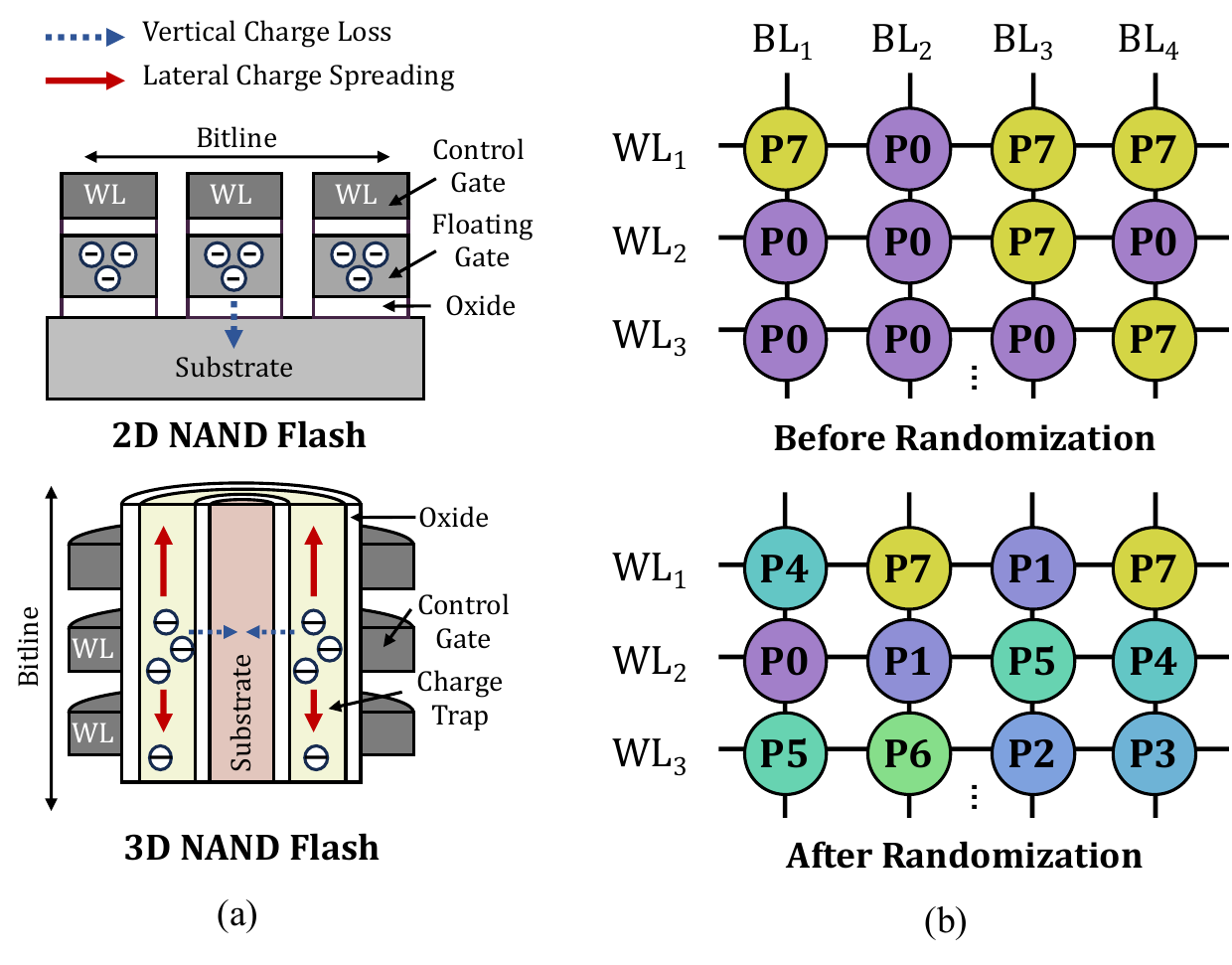}
  \captionof{figure}{(a) Mechanism of LCS. (b) Comparison of TLC data patterns before and after randomization}
  \vspace{-10pt}
  \label{fig:background2}
\end{figure}

A new type of flash cell (CT-type flash cell) is a key enabler of 3D flash memory to achieve cube-like 3D architecture because a conventional floating-gate flash cell was not suitable to stack flash cells along a vertical direction.
However, due to its unique structure and physics, 3D flash memory suffers from LCS, a new phenomenon in 3D flash memory.
Fig.~\ref{fig:background2}(a) illustrates the conceptual mechanism of LCS and why it is present only in 3D flash memory.
In 2D flash memory, since the floating gate of each cell is physically isolated, stored charges can leak only through the thin oxide layer between the floating-gate and substrate by various de-trapping mechanisms, which is called \emph{vertical charge loss}.
\jy{
3D flash memory, on the other hand, stores charges in a charge trap layer to increase the \vth.
Unlike the isolated floating gates in 2D flash memory, this charge trap layer is physically shared among adjacent cells along the bitline direction.
As a result, charges stored in one cell can unintentionally diffuse into neighboring cells, a phenomenon known as LCS.
Since LCS is driven by a diffusion process, the rate of charge migration is proportional to the concentration gradient of trapped charges between adjacent cells.
This means that greater differences in \vth levels between neighboring cells result in faster charge redistribution.
Because a cell’s \vth state is determined by its programmed data, LCS is highly sensitive to the data pattern across vertically neighboring cells.

}

\subsection{Data Randomizer}
In order to handle bit errors, it is common practice to employ strong error-correction codes (ECC) in modern SSDs.
ECC stores redundant bits called ECC parity, which enables detecting and correcting raw bit errors in the codeword. 
To cope with the high RBER of modern NAND flash memory, modern SSDs use sophisticated ECC that can correct several tens of raw bit errors per 1-KiB data (e.g., low-density parity-check (LDPC) codes~\cite{ldpc}).
However, when extremely unbalanced data patterns (e.g., all `0’ or `1' along the NAND string) are written to the NAND flash chip, shown in Fig.~\ref{fig:background2}(b), we may experience some unfortunate situations where RBER exceeds the maximum ECC capability~\cite{macronix}.
To avoid such events, commodity SSDs perform \emph{data randomization}, which scrambles user data before writing it to the NAND flash chips, thus forming a probabilistic random pattern (shown in Fig.~\ref{fig:background2}(b)).
The data randomizer transforms the original user data by simply inserting an exclusive OR (XOR) operation between the data path and the output of a Linear Feedback Shift Register (LFSR) initialized by a seed~\cite{randomizer}. 
%The seed is the initial value loaded into the LFSR to enable the random pattern generation.
The random seed is loaded into an internal circuit of the NAND Flash memory, called a page buffer, through the memory data-path. 
Then, additional peripheral circuits perform a bit-wise XOR operation on the original data input and a random seed in the page buffers, enabling the randomized write pattern.
When the randomized pattern is written, \vth states across different WLs are uniformly distributed, avoiding an unbalanced pattern. 
On the contrary, de-randomizing is performed during the read operation: after sensing the stored data from flash cells, the seed is loaded into a page buffer and a bit-wise XOR of the sensed data and random sequence is executed, reconstructing the original data.
Since the randomization algorithm exploits simple bitwise operations with a seed, its complexity is low, and both spatial and temporal overheads can be negligible.
These advantages make it widely adopted in commercial SSDs, where it effectively removes extremely unbalanced data.
}

\section{Motivation}
\label{sec:motivation}

\subsection{Impact Analysis of LCS in 3D Flash Memory}

\begin{figure}[t]
  \centering
  \includegraphics[width=\linewidth]{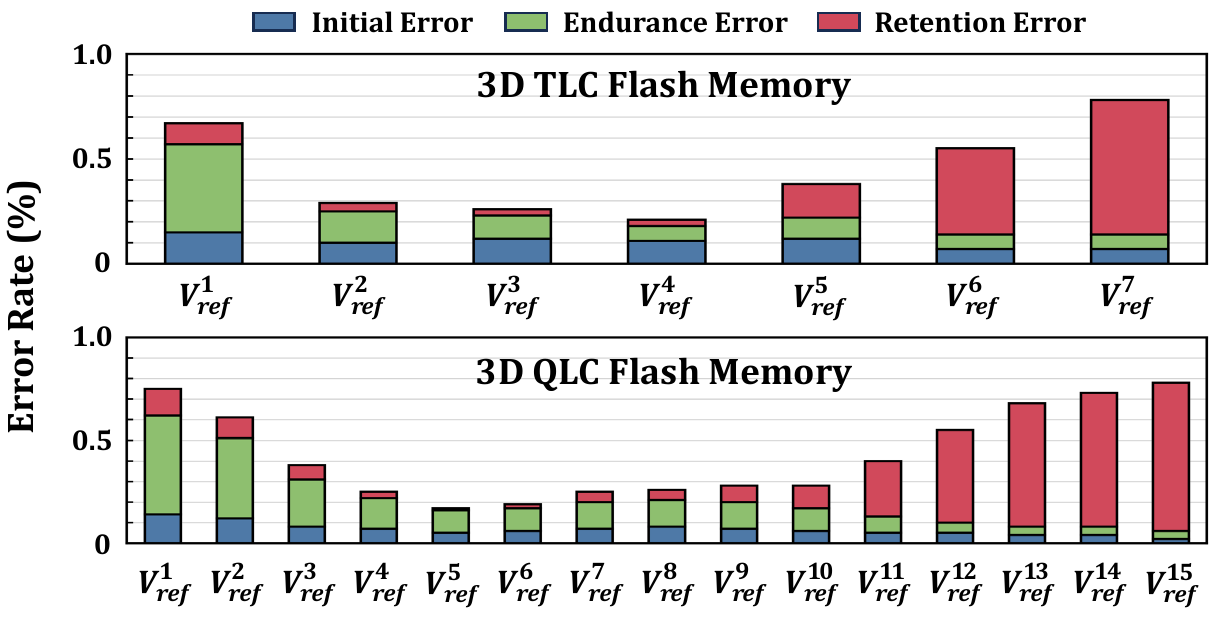}
  \caption{\jy{Comparison of inter-state error distributions between TLC flash memory (top) and QLC flash memory (bottom).}}
  \vspace{-10pt}
  \label{fig:state_error}
\end{figure}

To better understand the impact of LCS on the reliability of 3D NAND flash memory, we have conducted an extensive characterization study using 160 real 3D TLC/QLC flash chips (detailed test procedure is provided in Section ~\ref{sec:evaluation}).

\head{Inter-State Error Distribution} 
As is well known, reliability-related flash errors in low-capacity 2D flash memory are highly skewed across \vth states and occur in an asymmetric fashion~\cite{kim2018saro}. 
That is, not all bit errors occur with equal likelihood among different \vth states, and most bit errors are localized in a few specific \vth states. 
To quantify inter-state error distribution in 3D flash memory, we measured three types of errors in 3D TLC and QLC flash memory: initial (i.e., just after write), endurance, and retention errors.
Fig.~\ref{fig:state_error} shows the inter-state error distributions for TLC and QLC NAND flash memory, respectively.
From the evaluation results, we derive two key observations. First, the asymmetry in inter-state error distribution is more pronounced in QLC flash memory than in TLC flash memory. 
The error difference across \vth states is up to 3.7x in TLC flash memory, while it is aggravated up to 4.6x in QLC flash memory. 
Moreover, QLC flash memory exhibits twice as many error-prone \vth states as TLC flash memory, indicating that a greater variety of weak patterns may exist. 
Second, retention error is the dominant factor in determining the overall RBER in both TLC and QLC flash memory. Specifically, retention errors account for up to 48\% of total bit errors. 
Therefore, to ensure the reliability of high-capacity 3D flash memory, it is highly required to mitigate retention error by removing weak patterns.

\begin{figure}[t]
  \centering
  \setlength{\belowcaptionskip}{5pt}
  \includegraphics[width=\linewidth]{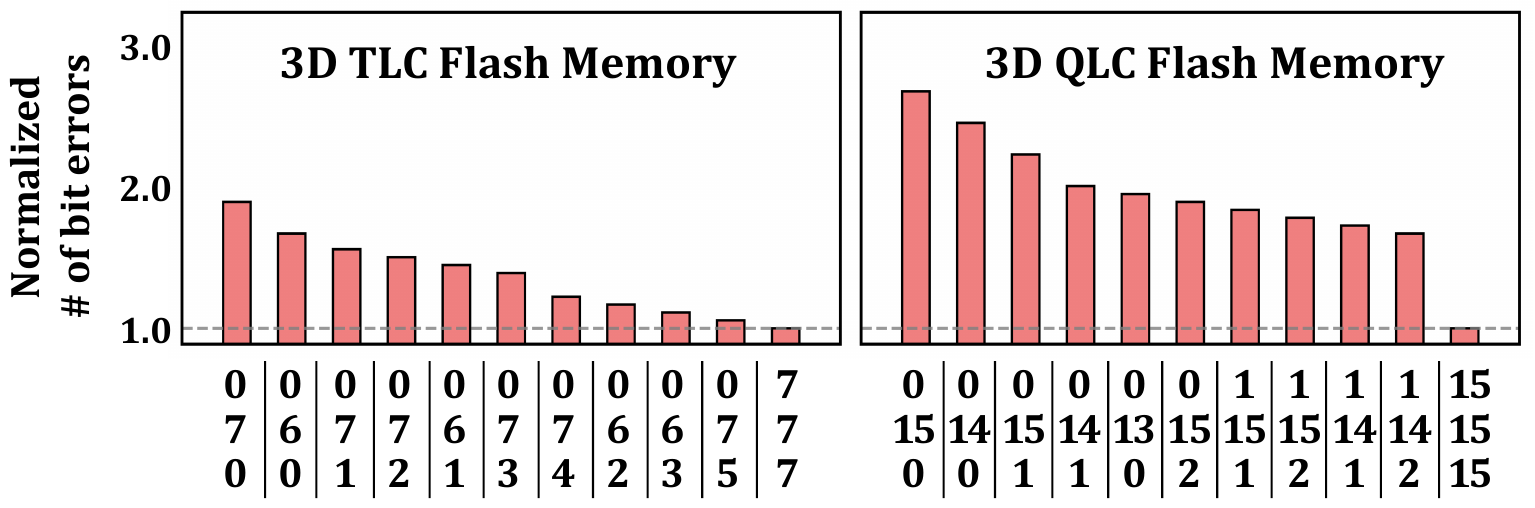}
  \caption{Top 10 LCS-induced weak patterns of TLC and QLC flash memory.}
  \vspace{-15pt}
  %left right는 불필요할 듯 하네요
  \label{fig:pattern}
\end{figure}

\head{LCS-Induced Weak Patterns}
As described in Section~\ref{subsec:back_LCS}, LCS occurs when charges diffuse into vertically adjacent cells through their shared charge trap layer, with diffusion rate proportional to the concentration gradient between neighboring cells. Consequently, patterns with large \vth differences between neighboring cells experience accelerated charge spreading, leading to higher LCS-induced bit errors. 
For example, in TLC NAND, if cells in the P7 state are adjacent to cells in the E state (i.e., an E–P7–E pattern), the victim cell in the middle (P7 state) is highly susceptible to errors due to LCS.
Similarly, in QLC NAND, a pattern such as E–P15–E can induce significant errors in the middle cell (P15 state).
To quantitatively understand the impact of \vth patterns on error, we measure the error rate of a victim cell under various \vth states of its adjacent cells.

Fig.~\ref{fig:pattern} shows the retention error rates of the top-10 LCS-induced weak patterns in TLC and QLC flash memory, ranked in descending order by their error contribution.
It demonstrates that patterns with larger \vth differences between the victim cell and vertically adjacent cells exhibit significantly higher bit error rates.
Notably, in TLC, the top-2 weak patterns produce substantially higher error rates than the remaining eight, whereas in QLC, the error contributions from the top-10 weak patterns are relatively comparable. This implies that in TLC, mitigating just few weak patterns is sufficient to reduce LCS-induced errors, while in QLC, it is necessary to suppress a broader range of weak patterns to achieve comparable error reduction. This difference can be attributed to the fact that QLC flash memory has twice as many states as TLC, resulting in approximately eight times more possible combinations of vertically adjacent patterns and forming a diverse set of LCS-induced weak patterns.

\subsection{Limitations of Existing Approaches}

Despite growing concern over LCS in modern 3D flash memory, how to effectively mitigate its reliability impact remains largely unexplored.
To our knowledge, only {\texttt{\small{TailCut}}}~\cite{lee2022tailcut} has proposed a specialized data encoding scheme aimed at mitigating LCS-induced weak patterns.
By extending existing peripheral circuitry within the flash chip, \tailcut effectively identifies and eliminates specific weak patterns in 3D TLC flash memory, achieving an average error reduction of 43.8\%.
However, \tailcut has two practical limitations that hinder its applicability to modern 3D NAND flash memory.

First, \tailcut requires modifications to the peripheral circuitry of existing flash chips, resulting in significant implementation complexity.
Due to limited hardware resources within flash chips, \tailcut is only capable of eliminating the top 2 weak patterns in TLC (E–P6–E and E–P7–E), which is insufficient for advanced flash technologies.

Second, \tailcut requires tracking bit information for all pages in the adjacent wordlines (WLs). For each programming operation, it must check whether $WL_{i-2}$ is in state E and $WL_{i-1}$ is in state P6/P7, then verify if $WL_i$ would be in state E. In high-capacity flash memory with four or more pages per cell (e.g., QLC or PLC flash memory), this approach introduces prohibitive overhead. The memory controller would need to perform multiple pattern checks to cover the expanded pattern space, significantly increasing both timing overhead and implementation complexity.

These limitations make \tailcut impractical for modern high-density NAND flash memory. To address reliability challenges in high-capacity QLC and next-generation PLC flash memory, a more generalized approach is needed—one that can efficiently handle diverse weak patterns without requiring flash chip modifications while maintaining compatibility with existing flash architectures.
\section{Overall Procedure of \STAR}
\label{sec:algorithm}

\begin{figure}[t]
  \centering
  \includegraphics[width=\linewidth]{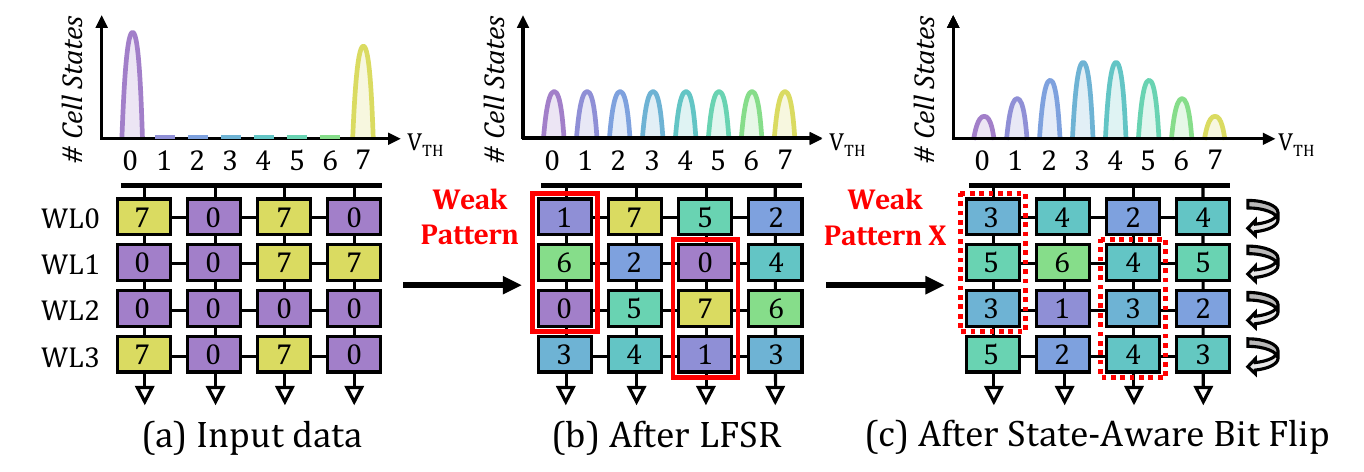}
  \caption{Illustrative example \STAR.}
  \vspace{-10pt}
  \label{fig:randomizer}
\end{figure}

\ms{
We propose \STAR \emph{(\stitle)}, a novel data randomizer scheme to improve SSD lifetime and performance by effectively mitigating the LCS-induced errors.
%reducing error-prone weak patterns.
Fig.~\ref{fig:randomizer} shows an illustrative example of how our proposed \STAR works.
The input data from the host can be highly biased, which can introduce unbalanced patterns that exceed the ECC capability (Fig.~\ref{fig:randomizer}(a)).
While modern SSDs can effectively prevent the unbalanced pattern using the LFSR-based data randomizer (Fig.~\ref{fig:randomizer}(b)), 
% uniform \vth distribution based on the data randomization introduces a new challenge to address the LCS problem. 
the resulting uniform \vth distribution is insufficient to address the new challenge posed by the LCS problem.

%cannot be the solution to the LCS problem.
%suffer from the LCS problem.
To overcome the challenge, \STAR performs an internal bit-flip operation on error-prone states so that it redistributes the number of each \vth state distribution to form a robust pattern for the LCS problem rather than a uniform data pattern (Fig.~\ref{fig:randomizer}(c)).
%By doing this, unlike existing techniques that target only specific worst-case patterns, \STAR effectively removes a wide spectrum of weak patterns, significantly improving the reliability against the LCS problem.
By doing this, unlike existing techniques that target only specific worst-case patterns, \STAR effectively removes a wide spectrum of weak patterns, significantly improving the reliability against the LCS problem.
%By simply modifying the conventional LFSR-based data randomizer, \STAR performs an internal bit-flip operation on the target error-prone states so that it redistributes the number of each \vth distribution to form a robust pattern to the LCS problem rather than a uniform data pattern.
%Unlike existing techniques that target only specific worst-case patterns, \STAR effectively removes a wide spectrum of weak patterns, significantly improving the reliability against the LCS problem.
%By doing this, unlike existing techniques that target only specific worst-case patterns, \STAR effectively removes a wide spectrum of weak patterns originating from the combination of error-prone states, significantly improving the reliability against the LCS problem.
To efficiently design \STAR, our technique proactively exploits a state-aware algorithm consisting of three sequential stages: \inum{i}~LFSR randomization, \inum{ii}~group error estimation, and \inum{iii}~optimal bit-flip. 
%To efficiently remove weak patterns, our technique proactively exploits a state-aware mechanism by incorporating three key steps. 

\head{LFSR Randomization}
The first stage in the \STAR exploits a conventional LFSR-based data randomizer to obtain a uniform \vth state distribution.
The LFSR-based data randomizer effectively breaks up inherent pattern dependencies in the original user data by scrambling it with random seed values, forming a probabilistically randomized pattern.
These randomized data are transferred into the next stage of \STAR as the input values to be bit-flipped, shown in Fig.~\ref{fig:algorithm} (\bcirc{1}).
%to the input data. This conventional randomization technique effectively breaks up any inherent pattern dependencies in the input data by distributing user data across wordlines (WLs) and bitlines (BLs) in a pseudo-random manner, as shown in Figure \ref{fig::randomizer}.

\head{Group Error Estimation}
Before performing the state-aware bit-flip operation, STAR estimates the error probability of each group based on the current state distribution of flash cells. In our technique \STAR, the error estimation is executed at the granularity of a group consisting of 128 flash cells (details on how to set a group are provided in Section~\ref{sec:implementation}). 
% Before performing the state-aware bit-flip operation, \STAR evaluates the impact of each bit-flip operation on the state distribution and error probability of each flash cell group.
% In our technique \STAR, the bit-flip operations and related functions are executed at the granularity of a group consisting of 128 flash cells (details on how to set a group are provided in Section V). 

\begin{figure}[t]
  \centering
  \includegraphics[width=\linewidth]{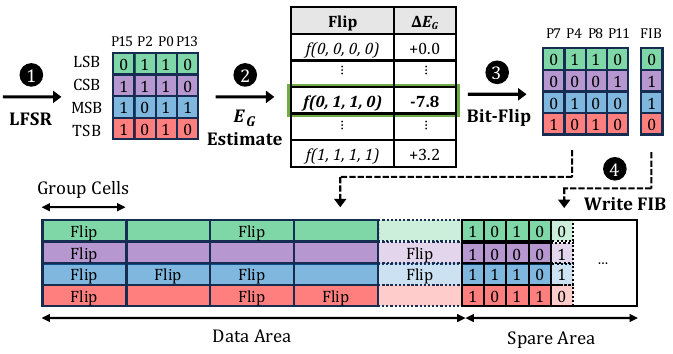}
  \caption{\STAR algorithm concept: (1) LFSR (2) Group Error Estimation (3) Optimal bit-flip (4) Write FIB.}
  \vspace{-10pt}
  \label{fig:algorithm}
\end{figure}

%This process is repeated for each group within the page, providing a fine-grained, group-by-group optimization. The following subsections detail each step of this process.
To quantify error probability at the group level, we define a new parameter, group error (E$_G$), as the cumulative error probability of all flash cells in the target group as follows:
%After LFSR-based randomization, \STAR evaluates the potential impact of various bit-flip operations on each group of cells. To quantify reliability at the group level, we define group error (E$_G$) as the cumulative error probability of all cells in a group:
\begin{equation}
E_G = \sum_{i=1}^{G} e_{{s_i}} =  \sum_{k=0}^{15} e_k N_k
\label{group_error}
\end{equation}
where:
\begin{itemize}
    \item \(G\) is the total number of cells in the group (e.g., 128),
%    \item \(G\) is the total number of cells in the group (e.g., \(G=128\)),
    \item \(s_i\) is the original state of the \(i\)-th cell in the group,
    \item \(k\) indexes the states from 0 to 15 (corresponding to P0 through P15 in QLC flash memory),
    \item \(e_k\) denotes the error probability of $k$-th state, 
    %\(P_k\), as determined by manufacturer-specific NAND 
    \item \(N_k\) is the number of cells in $k$-th state.
    %within the group.
\end{itemize}
%\STAR repeatedly performs the group error estimation for each group within the page.
%For example, for a 16-KiB page and group of 128 cells, \STAR calculates the \(E_G\) value 1,024 times for a bit-flip operation on a 16-KiB page of data.
A higher \(E_G\) value indicates that the group contains more error-prone states, resulting in a higher error probability. 
For calculating \(E_G\) value, \(e_k\) of each $k$-th state is determined using offline error profiling based on extensive real-device characterizations.

% STAR considers multiple combinations of bit inversions across the four bits stored in each QLC cell to explore how the state distribution can be optimized for error reduction.

% \STAR repeatedly performs the group error estimation for the target group until \(E_G\) is minimized through various bit-flip operations.
%Consequently, \STAR aims to reduce \(E_G\) by modifying \(\{N_0, \dots, N_{15}\}\) through optimal bit-flip operations.

%Based on the group error estimation, \STAR decides which bit in a group should be flipped, shown in Fig.\ref{fig:algorithm} (\bcirc{B}).
% Flipping specific bits in the current group reshapes the \vth state distributions of cells, which in turn changes the estimated value of \(E_G\). In our proposed \STAR, to represent bit-flip operation of which bit in a group is flipped, we introduce the notation as (assuming QLC flash memory):
Flipping specific bits in the current group reshapes the \vth state distributions of cells, which in turn changes the estimated value of \(E_G\). We refer to each specific combination of bit inversions as a bit-flip operation, and represent it using the following notation (assuming QLC flash memory):
%In QLC flash memory, since each flash cell stores four different bits, called LSB, CSB, MSB, and TSB bits, a bit-flip operation can selectively flip any subset of four bits. 
%In this work, we represent a bit-flip operation as:
\begin{equation}
f(b_{\text{TSB}}, b_{\text{MSB}}, b_{\text{CSB}}, b_{\text{LSB}})
\label{flip_operation}
\end{equation}
where \(b_j\in \{0, 1\}\) indicates whether the corresponding bit position is flipped (1) or not (0). 
For instance:
\begin{itemize}
    \item \(f(0, 0, 0, 0)\) does not change any bit in the data, 
    %(equivalent to an identity operation),
    \item \(f(0, 0, 0, 1)\) flips only the LSB bit in each flash cell,
    \item \(f(1, 1, 1, 1)\) flips all four bits (complete inversion).
\end{itemize}
In total, there are \(2^4=16\) different bit-flip operations. 
Each operation systematically transforms the state of each cell. 
For example, the bit-flip operation \(f(0, 0, 1, 0)\), which flips only the CSB, can convert state P0 (1111) to P9 (1101), or state P14 (0110) to P11 (0100).

To identify the impact of each bit-flip operation on the target group, \STAR calculates the difference between \(E_G\) values before and after for all possible bit-flip operations (Fig.~\ref{fig:algorithm} (\bcirc{2})).
%\footnote{Since calculating \(E_G\) values for all possible bit-flip operations is parallelized in \STAR, its impact on time overhead can be minimized. Detailed overhead analysis is provided in Section XX.)}. 
% Applying a bit-flip operation \(f\) to the target group transforms each cell’s original \vth state \(s_i\) into a new \vth state \(f(s_i)\), reshaping the overall \vth state distribuions.
%(e.g., 2$^4$=16 bit-flip operations).
When a bit-flip operation \(f\) is applied to the target group, the \(i\)-th cell in the group is converted from its original \vth state \(s_i\) to the new \vth state \(f(s_i)\).
%Such \vth state transformation across all cells in the group redistributes the overall \vth state distributions.
%Since the bit-flip operation changes the \(E_G\) value of the target group, the difference in the group error can be calculated as:
Therefore, the difference in the group error can be calculated as:
%%%%The altered state distribution resulting from bit-flip operations directly impacts the group error value. Since bit-flip operations act independently on each cell to produce these transformations, we can calculate the total change in group error as the sum of error changes at the individual cell level.
%%%%The difference in the group error is then:
\begin{equation}
\Delta E_{G,f} = \sum_{i=1}^{G} (e_{f(s_i)} - e_{s_i})
\label{error change}
\end{equation}
where:
\begin{itemize}
    \item \(s_i\) is the original state of the \(i\)-th cell in the group,
    \item \(f(s_i)\) is the transformed state after applying bit-flip \(f\),
    \item \(e_{s_i}\) and \(e_{f(s_i)}\) are the per-state error probability.
\end{itemize}
A negative \(\Delta E_{G, f}\) indicates that the bit-flip operation reduces the total error in the group, whereas a positive value indicates that it increases the error.
%In the next stage, we can find the optimal bit-flip operation that minimiz
By calculating the cumulative error change (\(\Delta E_{G, f}\)) for each bit-flip operation based on our characterized state-specific error rates, \STAR identifies which bit-flip operation would most effectively reduce the overall error probability of the group.

%Different bit-flip operations result in diverse redistributions of \vth states (i.e., changing the number of each \vth state).
%Therefore, by selecting the bit-flip operation having the largest negative \(\Delta E_{G, f}\), we can obtain the reduce the number of error-prone states, makin 
%Since different bit-flip operations result in diverse redistributions of \vth states (i.e., changing the number of error-prone states), by selecting the core objective is to identify the operation that most significantly decreases the group error.

\head{Optimal bit-flip}
%Based on the cumulative error change for each bit-flip option based on our characterized state-specific error rates, \STAR identifies which operation would most effectively reduce the group's overall error probability.
%Based on \(\Delta E_{G, f}\) value, we can find the optimal bit-flip operation that minimizes the RBER of the target group as follows:
%After computing \(\Delta E_{G, f}\) for all possible flips \(f\), \STAR selects the optimal flip:
\STAR selects the optimal bit-flip operation that minimizes the RBER of the target group as follows:
\begin{equation}
f^* = \arg\min_{f} \Delta E_{G,f}
\label{optimal_flip}
\end{equation}
% %%%Applying \(f^*\) to all cells in a group typically reduces that group’s overall error more effectively than any other groupwise bit-flip. The Flip Indicator Bit (FIB) which indicates whether the bit in that group were flipped) is recorded in the spare area to facilitate data recovery during subsequent reads. Moreover, because each group can exhibit a unique data distribution, applying an optimal flip on a per-group basis results in both finer control over error reduction and relatively low space overhead (since only the flip identifier must be stored). 
Since different bit-flip operations result in diverse redistributions of states (i.e., changing the number of each state), by selecting the bit-flip operation having the largest negative \(\Delta E_{G, f}\), we can reduce the number of error-prone states.
%Therefore, by selecting the bit-flip operation having the largest negative \(\Delta E_{G, f}\), we can reduce the number of error-prone states.
This approach systematically reshapes state distributions to favor lower-error states, statistically reducing the weak patterns without explicitly detecting them (Fig.~\ref{fig:algorithm} (\bcirc{3})). 

%These three stages in \STAR, LFSR randomization, group error estimation, and optimal bit-flip, are repeated for each group in a flash page, providing a fine-grained, group-by-group optimization.
These three stages in \STAR are repeated for each group in a flash page, providing a fine-grained, group-based optimization.
The optimal bit-flip information for each group is recorded in the Flip Indicator Bit (FIB) within the spare area of the target flash page (Fig.~\ref{fig:algorithm} (\bcirc{4})).   
FIB exploits a single bit for each of the LSB, CSB, MSB, and TSB bit positions within the group, where `1' indicates the bit was flipped and `0' indicates it was not flipped.
%FIB is stored in the spare area of the flash page during a NAND program operation.
In subsequent read operations for corresponding flash pages, the FIB information is used to reverse the bit-flip operations.
Then, the conventional de-randomization is performed to fully recover the original data.
%before performing de-randomization, ensuring data recovery.
%This process is repeated for each group within the page, providing a fine-grained, group-by-group optimization. The following subsections detail each step of this process.
%\STAR repeatedly performs the group error estimation for each group within the page.
%For example, for a 16-KiB page and group of 128 cells, \STAR calculates the \(E_G\) value 1,024 times for a bit-flip operation on a 16-KiB page of data.

%Applying \(f^*\) to all cells in a group reduces the group error the most. Since optimization is performed at the group level, different optimal bit-flip operations can be selected for each group as shown in Figure \ref{fig::algorithm}. This approach allows data to be randomized in a way that minimizes error according to the unique data distribution of each group. 

%As illustrated in Fig \ref{fig::algorithm}, Information of optimal bit-flip operation selected for each group is recorded through a Flip Indicator Bit (FIB). 
%This metadata uses a single bit for each of the LSB, CSB, MSB, and TSB positions within the group, where 1 indicates the bit was flipped and 0 indicates it was not flipped. FIB is stored in the spare area of each flash page. During read operations, the FIB is used to recover the flipped data by flipping each bit again if it indicates 1. Then LFSR derandomization is performed to fully recover the original data before transmission to the host.

}

\section{Design and Implementation of \STAR}
\label{sec:implementation}

\begin{figure*}[t]
  \centering
  \captionsetup[subfigure]{skip=2pt}
  \setlength{\belowcaptionskip}{2pt}
     \includegraphics[width=\linewidth]{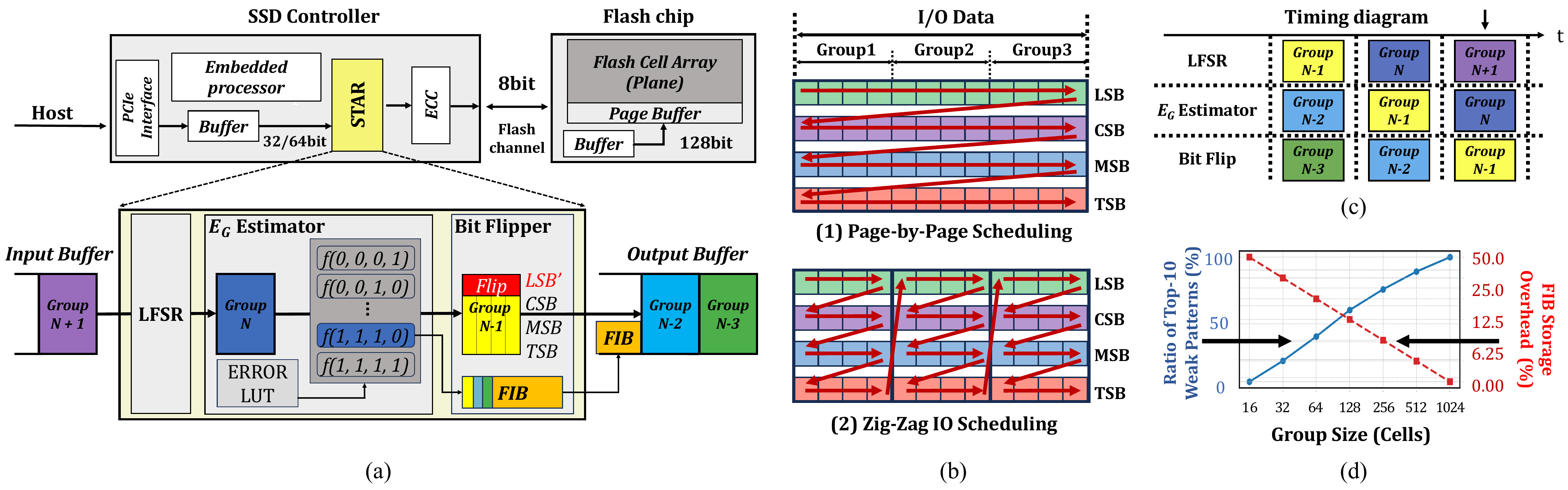}
  \captionof{figure}{(a) Hardware Design of \STAR. (b) Zig-Zag IO Scheduling. (c) Group-Level Pipelining. (d) Group Size Selection.}
  \vspace{-10pt}
  \label{fig:design_implementation}
\end{figure*}

The \STAR algorithm is rather straightforward to implement, but a naive implementation can incur significant capacity and performance overheads. 
In this section, we present our design choices to meet the key implementation requirements of \STAR.

\subsection{Design Optimization}
As illustrated in Fig.~\ref{fig:design_implementation}(a), \STAR is implemented within the data path of the SSD controller between the host interface and the NAND flash chip.
It consists of three modules (LFSR, \(E_G\) Estimator, and Bit-Flipper) which correspond to key stages described in Fig.~\ref{fig:algorithm}.
To fully leverage its reliability benefits while minimizing overhead, \STAR incorporate three optimizations: (1) Zig-Zag IO Scheduling, (2) Group-Level Pipelining, and (3) Parallel Error Estimator, each of which is enabled by efficient hardware integration.

\head{Zig-Zag IO Scheduling (ZIS)}
As explained in Section IV, STAR performs state-aware bit-flip operations repeatedly for each group, which is a time-consuming process requiring a huge amount of computation. For example, in QLC flash memory with a 16-KiB page and 128-cell groups, STAR needs to calculate (4096 $\times$ 16) \(E_G\) values to find the optimal bit-flip operation on one WL data.

In modern SSDs, user data passes through the SSD controller in page order (i.e., page-by-page scheduling) as shown in Fig.~\ref{fig:design_implementation}(b).
It means that all LSB data is sent to the flash chip first, followed by CSB, MSB, and TSB in a sequential manner.
Since \STAR operates on a basis of groups, the conventional page-by-page scheduling makes it impossible to integrate the bit-flip operation into the data I/O path.

Therefore, to address the challenge, we introduce a new optimization, Zig-Zag IO scheduling.
It ensures that each bit of LSB, CSB, MSB, and TSB is transferred to the SSD controller with interleaving at a group granularity, allowing all four bits within a group to pass through the SSD controller simultaneously.
Since \STAR is designed to operate at a throughput of 32 or 64 bits per cycle, equal to the throughput of external buffer connected to \STAR, only the initial latency of a single group passing through \STAR is added to the overall I/O performance.
100 ns of latency will not act as a bottleneck in the I/O path, with no impact on the overall data I/O throughput.

\head{Group-Level Pipelining (GLP)}
Although all four bits in a group can be processed simultaneously due to the optimized I/O scheduling, the I/O throughput can be severely degraded as the operations for each group proceed sequentially.
To overcome this problem, we design an optimization technique, Group-Level Pipelining, by interleaving the per-group computational tasks within the \STAR. 
As shown in Fig.~\ref{fig:design_implementation}(c), due to three dedicated hardware modules for LFSR, \(E_G\) Estimator, and Bit-Flipper, different operations of each group can be performed simultaneously.
For example, while one group is under the Group Error Estimation module, the previous group simultaneously proceeds to the Bit-Flipper module, and the next group enters the LFSR Randomization module.
By implementing micro-pipelining that reflects the different latencies of each module, we can match the number of groups processed by \STAR per cycle and the input rate of groups from the external buffer.
This carefully designed overlapping ensures continuous data flow without bottlenecks in the I/O path, allowing \STAR to achieve high data I/O throughput.

\head{Parallel Error Estimator (PEE)}
Even under the GLP processing, the long-latency module can significantly reduce the pipelining efficiency, which can bottleneck the overall performance of \STAR. 
Since the \(E_G\) estimator is the slowest module, consuming the majority of the processing time, we optimize the overall procedure for the \(E_G\) estimator through a dedicated hardware architecture.
Our goal is to evaluate the impact of all 16 possible bit-flip operations on group error (\(E_G\)) simultaneously.

To this end, we implement \(E_G\) estimator with 16 dedicated computation units, each responsible for calculating the cumulative error change for a specific bit-flip operation.
Each unit linearly scans the cells in a group, accumulating the error difference between original and transformed states as defined in Eq.~\ref{error change}.
To efficiently calculate the cumulative error change for all possible state transitions (from original state to transformed state), the \(E_G\) estimator exploits pre-characterized error-change values, which are stored in an on-chip Look-Up Table (LUT). 
% Each value in the LUT is derived from offline error analysis, based on a thorough characterization of real 3D flash memory.
Each value in the LUT is derived from offline error analysis of real 3D flash memory.
During runtime operation, each computation unit fetches the required error-change values from the LUT, thereby eliminating the need for complex calculations.
This allows \STAR to determine the optimal bit-flip operation with minimal delay while maintaining full I/O throughput.

\subsection{Group Size Selection}
The group size is a critical factor in deciding the efficiency of \STAR.
As shown in Fig.~\ref{fig:design_implementation}(d), since the group size affects both how much weak patterns can be reduced and how much space is needed for FIB overhead, it should be set carefully to maximize the benefit of our technique.
For example, a smaller group size enables more fine-grained control over bit-flip optimization, which results in better error reduction by removing more weak patterns.
However, when group size decreases, the number of groups per page increases proportionally, requiring more FIB bits to store the bit-flip information (i.e., need for more spare area).
% Additionally, FIB is recommended to fit within 200B per page in typical QLC flash memory configurations, as the flash spare area should support ECC parity bits and other system metadata~\cite{li-micro-2020}.
% As a result, in our \STAR, we select a 128-cell group to balance between various constraints, such as error reduction, FIB overhead, and allowable flash spare area.
\okj{
Recent work~\cite{li-micro-2020} shows that reserving up to 1\% of cells per wordline cells in the space area for auxiliary metadata has a negligible impact on ECC correction capability. Accordingly, we select a 128-cell group size in our QLC flash design (16KB page, 2KB spare area), resulting in a FIB overhead of only 0.7\%, well within the practical limitation, while maximizing error reduction.
}

\subsection{Analysis of Implementation Overhead}
\STAR requires only minor changes to existing SSDs, without modifying NAND chips.
To evaluate the overhead, we implemented the \STAR hardware logic in HDL and conducted \okj{Power, Performance, and Area (PPA)} analysis using Synopsys Design Compiler, targeting a 45nm standard-cell library.
First, the performance overhead of \STAR is minimized through latency-efficient design.
\STAR only adds less than 100 ns to the conventional data randomization process, which is negligible compared to typical QLC flash memory program operations that range from tens to thousands of microseconds, ensuring no degradation in IOPS performance in practice.
Second, chip-area overhead is also negligible in practice due to a compact hardware design. 
The analysis based on EDA tool demonstrates that implemented in a 45nm process, the \STAR randomizer occupies only 0.036mm$^{2}$ (44.8K gates), representing a negligible fraction compared to a typical SSD controller die size of approximately 70mm$^{2}$~\cite{tavakkol-fast-2018, cornwell2012anatomy}.
Third, the power analysis shows that \STAR incurs additional power consumption of approximately 15.3mW. 
Since it accounts for less than 0.3\% of the overall power budget in typical SSDs (5W–15W)~\cite{techpowerup-2024}, the power overhead can be trivial.
\okj{
Finally, \STAR stores a single FIB bit per group in the flash spare area. 
With a 128-cell group size, the FIB overhead is only 0.7\% of the wordline, which is well within practical limits~\cite{li-micro-2020}.
}
\section{Evaluation} 
\label{sec:evaluation}

\begin{figure}[t]
  \centering
  \captionsetup[subfigure]{skip=2pt}
  \setlength{\belowcaptionskip}{2pt}

  \subfloat[Distribution of each \vth state in TLC/QLC NAND.\label{fig:state_distribution}]{
    \includegraphics[width=\linewidth]{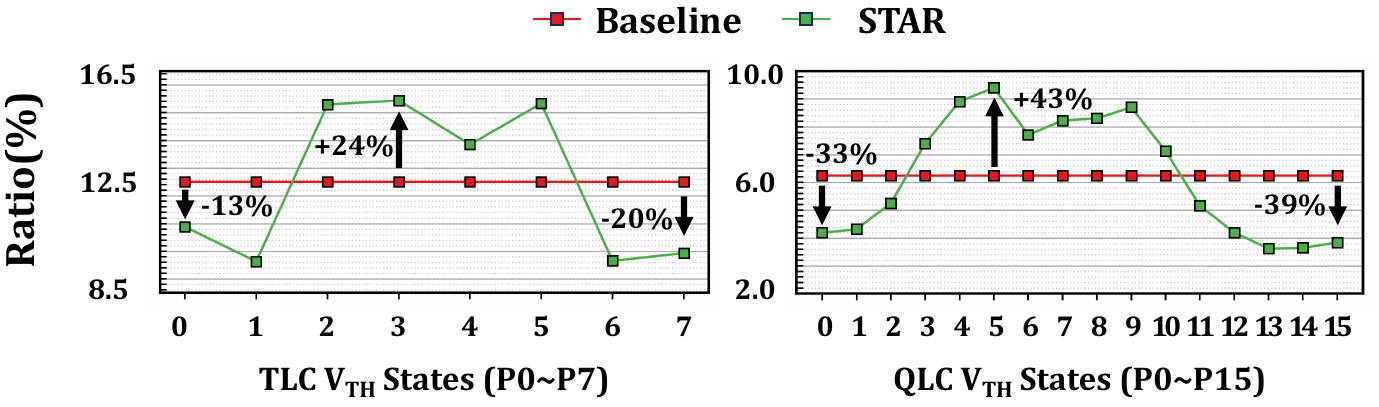}
  }\\[2pt]
  \subfloat[Suppression of the top 10 weak patterns in TLC/QLC NAND.\label{fig:weak_pattern}]{
    \includegraphics[width=\linewidth]{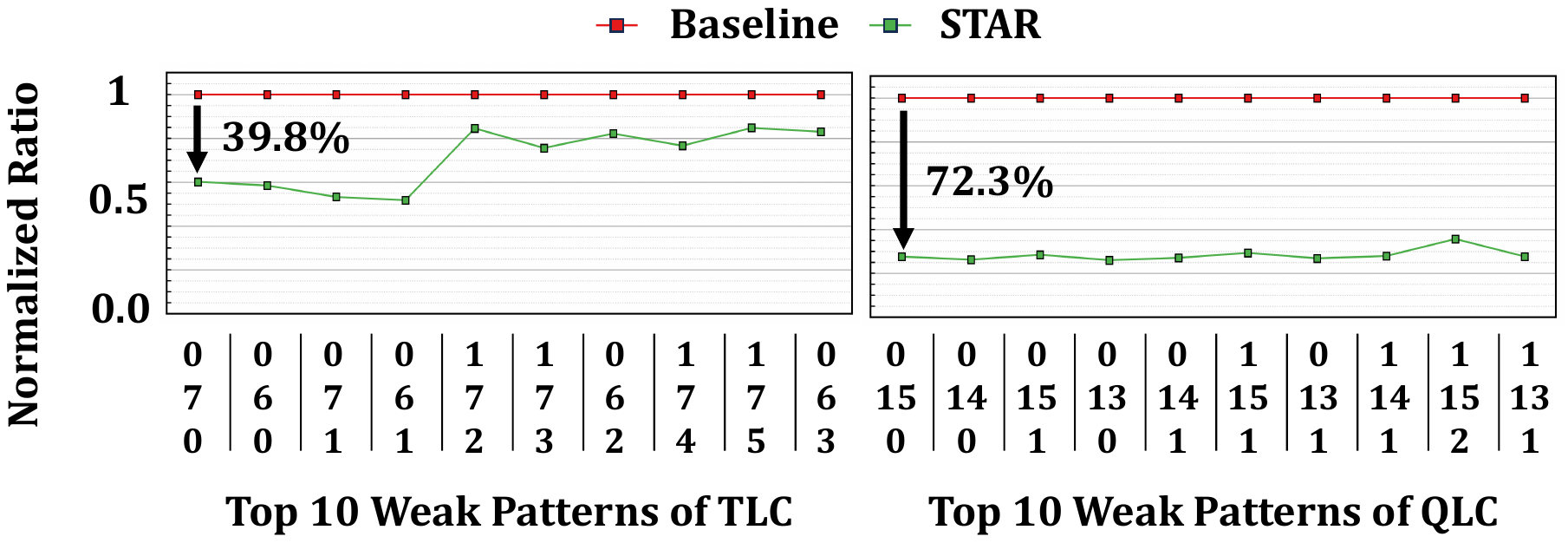}
  }

  \caption{Impact of \STAR on 3D TLC/QLC NAND: (a) \vth state distribution and (b) normalized suppression of weak patterns.}
  \vspace{-5pt}
  \label{fig:STAR_simulated}
\end{figure}

% To evaluate the effectiveness of \STAR, we implemented a \STAR-aware SSD, \StarSSD, using a state-of-the-art NVMe-Virt emulator~\cite{kim2023virt}.
% We extended NVMeVirt to emulate reliability characteristics by modifying its internal NAND flash model.
% Key parameters in our evaluation are extracted through real-device characterizations using 160 real 3D TLC/QLC flash chips.
% To minimize the potential distortions in our results, we evenly select 120 blocks from each chip at different physical locations and test all WLs (a total of 3,686,400 WLs) in each selected block.
% By using an FPGA-based testing platform with a custom flash controller and a temperature controller, we measured flash characteristics (e.g., endurance error, retention error, and read-retry counts) while varying both the number of P/E cycles (denoted as PEC) and data retention time. 
% Each test scenario follows the JEDEC standard~\cite{jedec} common in the memory industry.
% For example, to emulate a 12-month retention time condition, we baked the flash chips at 85\degree for 13 hours (which is equivalent to one year at 30\degree based on the Arrhenius’s law~\cite{arr}).

To evaluate the effectiveness of \STAR in real hardware, we conducted extensive device-level characterization using 160 commercial 3D TLC/QLC flash chips. 
We implemented three SSD configurations: one equipped with a conventional LFSR-based randomizer (Baseline), one with TailCut, and one with our proposed \STAR randomizer.
For each chip, we uniformly selected 120 blocks across different physical locations and tested all wordlines—amounting to 3,686,400 WLs in total.
Using an FPGA-based testing platform with a custom flash controller and a temperature chamber, we measured key reliability metrics—endurance errors, retention errors, and read-retry counts—under varying program/erase (P/E) cycles and data retention times.
Test protocols followed JEDEC industry standards~\cite{jedec}; for example, to simulate a 12-month retention period, chips were baked at 85\degree for 13 hours, equivalent to one year at 30\degree based on Arrhenius’s law~\cite{arr}.

Using a state-of-the-art SSD emulator~\cite{kim2023virt}, we construct a \STAR-aware SSD model, \StarSSD.
Specifically, we extend the NAND model to reflect LCS error behaviors and read-retry dynamics based on our real-chip characterization.
The emulator includes three SSD configurations—Baseline (LFSR), TailCut, and \StarSSD—allowing system-level evaluation under realistic error conditions. As summarized in Table~\ref{tab:ssd_config_latency}, we configured the architecture and key parameters of the emulated SSDs to be close to modern high-end SSDs employing state-of-the-art 3D flash memory~\cite{samsung870evo,samsung870qvo}.
\begin{table}[b]
\scriptsize
\caption{Key parameters of a target SSD~\cite{samsung870evo, samsung870qvo}.}
\label{tab:ssd_config_latency}
\renewcommand{\arraystretch}{1.1}
\begin{tabularx}{\linewidth}{@{} >{\raggedright\arraybackslash}p{2.8cm} | X @{}}
\toprule
\textbf{Configuration} & 1\,TB total capacity,\quad 8 channels, \\
                       & 1 chips/channel,\quad 2 planes/die \\
\midrule
\textbf{Interface}     & NVMe 8\,GB/s \\
\midrule
\boldmath$t_\mathrm{R}$ (\textmu s)\unboldmath & 45 (TLC), 110 (QLC) \\
\midrule
\boldmath$t_\mathrm{PROG}$ (\textmu s)\unboldmath & 390 (TLC), 2000 (QLC) \\
\bottomrule
\end{tabularx}
\end{table}
 For our evaluations, we study five workloads from the Filebench benchmark suite: OLTP, File, Proxy, Mail, and Web server~\cite{filebench}. 
Table~\ref{tab:filebench_workloads} summarizes the I/O characteristics of the workloads used for our evaluations~\cite{filebench-workload}.

\captionsetup[table]{skip=4pt}
\begin{table}[b]
\centering
\tiny
\caption{Key I/O characteristics of evaluated workloads~\cite{filebench-workload}.}
\renewcommand{\arraystretch}{0.9}
\resizebox{1.0\columnwidth}{!}{
\begin{tabular}{c|c c c c}
\toprule
\textbf{Workload} & 
\textbf{File size} & 
\textbf{\# files} & 
\textbf{\# threads} & 
\textbf{R/W Ratio} \\
\midrule
Web Server  & 32 KB    & 20,000  & 100       & 10:1 \\
File Server & 256 KB   & 50,000  & 100       & 1:2  \\
Mail Server & 16 KB    & 50,000  & 100       & 1:1  \\
DB Server   & 0.5 GB   & 10      & 200+10    & 20:1 \\
Web Proxy   & 1 MB     & 10,000  & 100       & 5:1  \\
\bottomrule
\end{tabular}
}
\label{tab:filebench_workloads}
\end{table}

\label{evaluation}

\subsection{Impact on SSD Lifetime}

\begin{figure}[t]
  \centering
  \captionsetup[subfigure]{skip=2pt}
  \setlength{\belowcaptionskip}{2pt}
  \subfloat[Normalized lifetime comparison across chip percentiles (Best to Worst).\label{fig:lifetime}]{
    \raisebox{-2pt}{\includegraphics[width=\linewidth]{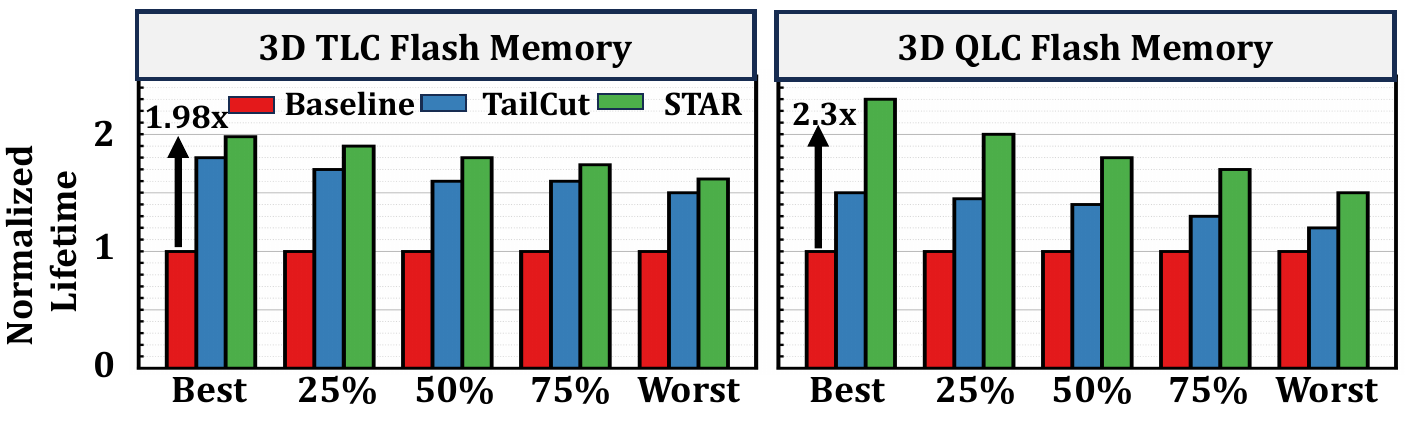}}
  }
  \\[2pt]
  \subfloat[Read retry counts across varying retention times and P/E cycles.\label{fig:read_retry_counts}]{
  \includegraphics[width=\linewidth]{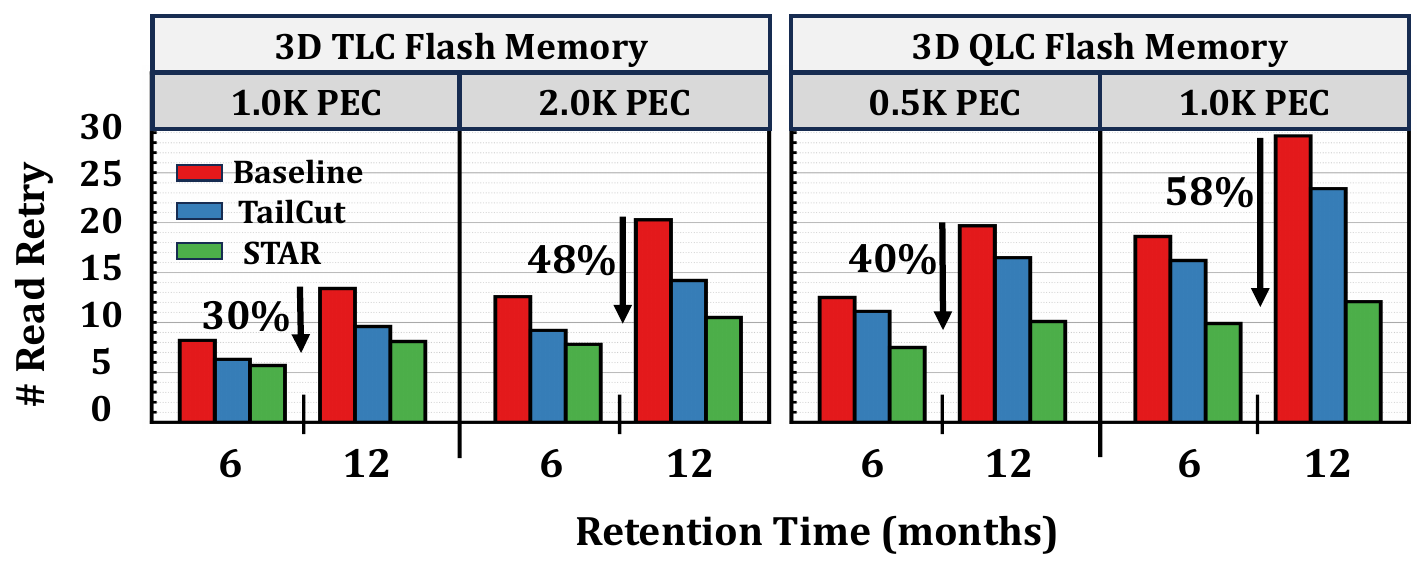}
}
  \caption{Impact of \STAR on (a) lifetime and (b) read retry counts, measured using 160 real 3D TLC/QLC flash chips.}
  \vspace{-10pt}
  \label{fig:STAR_realdevice}
\end{figure}

% \begin{figure*}[t]
%   \centering
%     \setlength{\belowcaptionskip}{1pt}
%     \includegraphics[width=\linewidth]{graphs/ssd_response.pdf}
%   \caption{Comparison of response time of (a) TLC and (b) QLC on Filebench workloads.}
%   \vspace{-10pt}
%   \label{fig:ssd_response}
% \end{figure*}

\begin{figure*}[t]
  \centering
  \setlength{\belowcaptionskip}{1pt}
  \includegraphics[width=0.8\linewidth, height=0.25\textheight, keepaspectratio]
  {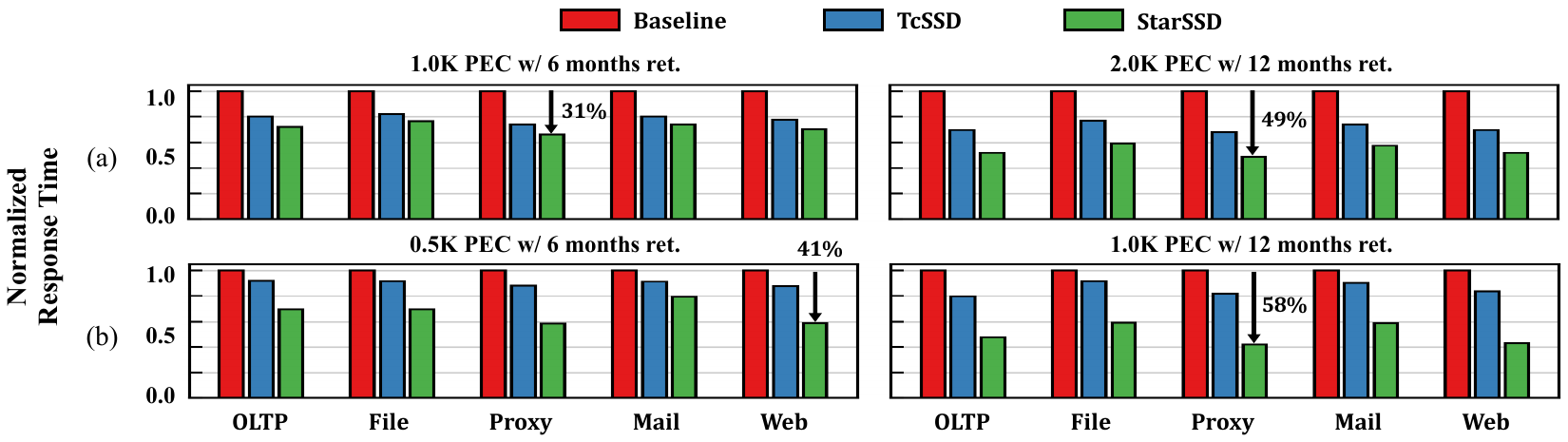}
  \caption{Comparison of response time of (a) TLC and (b) QLC on Filebench workloads.}
  \vspace{-10pt}
  \label{fig:ssd_response}
\end{figure*}

\head{Reduction of Weak Patterns}
As shown in Fig.~\ref{fig:STAR_simulated}(a), \STAR reduces weak patterns by reshaping the \vth state distributions.
In TLC flash memory, \STAR lowers the error-prone states P0 and P7 by 13\% and 20\%, respectively, compared to baseline.
The effect is even more pronounced in QLC flash memory, where \STAR reduces the P0 and P1 by 33\% and 31\%, and P14 and P15 by 42\% and 39\%, respectively.
This greater reduction in QLC flash memory is due to \STAR’s ability to exploit more bit-flip candidates—4 bits per cell versus 3 in TLC—enabling more effective bit-flip pattern optimization.
The reduced error-prone states lower weak pattern generation, as shown in Fig.~\ref{fig:STAR_simulated}(b).
% In TLC flash memory, \STAR reduces the E-P7-E pattern (the top weak pattern) by 39.8\% in TLC flash memory and decreases the top 10 weak patterns by an average of 29.0\%.
In TLC flash memory, \STAR reduces the top weak pattern E-P7-E by 39.8\% and reduces the occurrence of the top 10 weak patterns by an average of 29.0\%.
\STAR is more effective in QLC flash memory, reducing the E-P15-E pattern by 72.3\% and the top 10 weak patterns by an average of 71.7\% over the baseline.

\head{Lifetime Improvement}
\STAR improves SSD lifetime by reducing the RBER via eliminating weak patterns.
To quantify this improvement, we subjected devices to increasing P/E cycles and identified the point at which the RBER exceeded the ECC correction capability (72 bits per 1-KiB)~\cite{macronix2021ecc} for each comparison scheme.
Fig.~\ref{fig:STAR_realdevice}(a) shows that \STAR significantly extends SSD lifetime. For TLC flash memory, \STAR increased lifetime from 7.5K PEC to 9.9K PEC (98\%), compared to \tailcut's 7.5K PEC to 9.0K PEC (80\%). 
For QLC flash memory, \STAR increased lifetime from 1.5K to 2.3K PEC (130\%), compared to \tailcut’s 1.2K to 1.5K PEC (50\%), highlighting the greater effectiveness of \STAR in QLC flash memory. 
Unlike \tailcut, which targets specific weak patterns, \STAR aims to redistribute error-prone states, significantly reducing a majority of weak patterns.
These results indicate that \STAR is especially effective for flash memories with many error-prone states and weak patterns, such as QLC flash memory.

\subsection{Impact on Read Performance}

\head{Reduction in Read Retry Counts}
By minimizing weak patterns, \STAR effectively reduces unintended \vth shifts caused by LCS, thereby lowering the frequency of read retry, which is one of the primary contributors to read latency in modern 3D flash memory~\cite{chun-hpca,lee2022tailcut}.
Fig.~\ref{fig:STAR_realdevice}(b) shows the average read-retry count under varying PEC and retention times.
For TLC flash memory, \STAR reduced read-retry operations by up to 48\% after 2.0K PEC compared to the baseline. QLC flash memory showed even greater improvements: up to 40\% and 58\% reduction after 0.5K and 1.0K PEC, respectively, implying a high probability of SSD latency improvement.

\head{Read Latency Improvement}
To assess read performance, we compare \StarSSD with a \tailcut-based SSD (\TcSSD) using the read-retry models in Fig.~\ref{fig:STAR_realdevice}(b). 
Average read latency per operation was measured across five workloads (Tab.~\ref{tab:filebench_workloads}). Fig.~\ref{fig:ssd_response} shows normalized read response times of \StarSSD and \TcSSD against the \baseline under various PEC and retention settings. In TLC SSDs (Fig.~\ref{fig:ssd_response}(a)), \StarSSD reduced latency by 28\% over the \baseline and 9\% over \TcSSD at 1.0K PEC with 6-month retention. With further aging (2.0K PEC, 12-month retention), reductions reached 46\% and 25\%, respectively. For QLC SSDs (Fig.~\ref{fig:ssd_response}(b)), where RBER and retention errors are more severe, the improvement is more pronounced. At 0.5K PEC and 6 months, \StarSSD reduced latency by 33\% over the \baseline and 25\% over \TcSSD; at 1K PEC and 12 months, the reduction reached 50\% and 41\%, respectively. Although \TcSSD also improves over the \baseline, its gains in QLC are limited due to diverse weak patterns. \StarSSD consistently outperforms both baselines across all workloads and NAND types, demonstrating strong scalability and robustness under varied conditions.

\section{Related Work}
To our knowledge, this work is the first to address the LCS problem by redesigning the data randomizer in the SSD controller, providing significant lifetime and performance benefits for modern SSDs without modification to the NAND flash chip. 

\head{LCS Problem}
Several studies have characterized the mechanism of LCS in 3D NAND flash memory. Mizoguchi et al.\cite{mizoguchi} revealed the physical mechanism behind LCS by investigating the amount of \vth shift in 3D charge-trap cells, while Luo et al.~\cite{luo-acm-2018} demonstrated that LCS contributes up to 80\% of early charge loss in 3D flash memory. However, these works focused only on device-level characterization without analyzing the impact of LCS on the performance and reliability at the system level.
The recent system-level approach, \tailcut~\cite{lee2022tailcut}, improves the LCS problem by detecting and eliminating the two most vulnerable weak patterns.
However, it requires flash chip modifications with complex read/write operations, limiting its ability to eliminate only a few patterns and reducing effectiveness in advanced flash technologies with many weak patterns.

\head{Randomizer}
Conventional data randomization is widely used in modern SSDs~\cite{randomizer} to prevent unbalanced patterns from exceeding ECC capability. However, existing data randomizers do not address the challenges posed by LCS. To overcome this, \STAR employs a state-aware bit-flip mechanism to reshape \vth states and improve robustness against LCS.
\section{Conclusion}
In this paper, we introduced \STAR, a novel randomization technique addressing reliability challenges in 3D NAND flash memory. 
Unlike conventional LFSR-based randomizers which uniformly distribute data across \vth states, \STAR strategically reshapes \vth states by applying optimal group-level bit-flip operations, effectively reducing error-prone states and minimizing LCS effects.
By lowering error-prone states (e.g., P0, P1, P14, P15), \STAR statistically reduces weak patterns without explicitly detecting them.
Our comprehensive evaluations demonstrate significant improvements, extending the lifetime of TLC and QLC flash memory by up to 1.98x and 2.3x compared to baseline, and improving read performance of TLC and QLC-based SSD by an average of 46\% and 50\% on real-world traces.
Importantly, \STAR's controller-level implementation requires no NAND flash chip modifications and adds minimal overhead, making it highly practical.
Although tested on TLC and QLC flash memory, \STAR’s design is expected to be effective in next-generation SSDs such as PLC flash memory, which pose more severe reliability challenges.
\section{Acknowledgment}
This work was mainly supported by Korea Institute for Advancement of Technology(KIAT) grant funded by the Korea Government (Ministry of Education) (P0025681-G02P22450002201-10054408, Semiconductor - Specialized University), by the Ministry of Science and ICT (MSIT, Korea) under Grant RS-2024-00456287, and by the Institute for Information \& communications Technology Planning \& evaluation (IITP) under Grants RS-2024-00459026 and RS-2025-02214654.
Myungsuk Kim was supported in part by the National Research Foundation of Korea (RS-2024-
00414964) and by K-CHIPS(Korea Collaborative \& High-tech Initiative for Prospective Semiconductor Research) (2410012261, 02219428, 25066-15FC) funded by the Ministry of Trade, Industry \& Energy (MOTIE, Korea).
The ICT at Seoul National University provided research facilities for this study.
The EDA tool was supported by the IC Design Education Center (IDEC), Korea.
\emph{(Co-corresponding Authors: Jihong Kim and Myungsuk Kim)}

%%%%%%%%% -- BIB STYLE AND FILE -- %%%%%%%%
%\bibliographystyle{unsrt}
\bibliographystyle{IEEEtran}
\bibliography{refs}

\end{document}